# Single-mode InAs/GaAs quantum-dot DFB laser with oxidized aperture confined surface grating

*Zhengqing Ding[1], Anyao Zhu[1], Chaoyuan Yang[1], Kun Zhan[1], Yingxin Chen[1], Ying Yu[1]\*, Siyuan Yu[1]*

[1] State Key Laboratory of Optoelectronic Materials and Technologies, School of Electronics and Information Technology, Sun Yat-Sen University, Guangzhou 510275, China

**Abstract**
InAs/GaAs quantum dot (QD) distributed feedback (DFB) lasers are promising candidates for next-generation photonic integrated circuits. We present a design that incorporates an oxidized aperture confined surface grating (OASG) structure, which reduces non-radiative recombination losses and surface optical losses sustained in device fabricated by conventionally fabrication methods including etching and regrowth. The OASG-DFB laser eliminates the need for ridge waveguide etching and avoids instability in sidewall grating coupling. Experimental results show stable single-mode operation, a maximum output power of 15.1 mW, a side-mode suppression ratio (SMSR) of 44 dB, and a narrow linewidth of 1.79 MHz. This approach simplifies fabrication, reduces costs, and enhances the scalability of GaAs-based QD DFB lasers for applications in optical communication and photonic integration.

Self-assembled InAs/GaAs quantum dots (QDs), with their unique three-dimensional carrier confinement, exhibit superior properties in semiconductor lasers, including low threshold current density[1], high temperature stability[2], low relative intensity noise (RIN)[3–5], and high tolerance to external feedback[6–9], making them ideal candidates for next-generation photonic integrated circuit (PIC) light sources[10]. Despite these advantages, fabricating high-performance InAs/GaAs QD distributed feedback (DFB) lasers faces several significant challenges.

Traditional fabrication approaches for GaAs-based QD DFB lasers often follow similar procedures as those used for InP-based devices[11], where a grating structure is etched directly on the surface of the wafer, and followed by a regrowth process to form the cladding and contact layers. The regrowth process typically involves molecular beam epitaxy (MBE) for AlGaAs cladding layers[12] or metal-organic chemical vapor deposition (MOCVD) for AlGaAs layers[13] and InGaP layers[14]. However, these approaches encounter several issues: the oxidation of Al-containing materials during regrowth can degrade material quality, and although InGaP growth occurs at lower temperatures than AlGaAs, reducing thermal damage to QDs, the MOCVD regrowth of InGaP is susceptible to phase separation, potentially compromising device performance.

To overcome these challenges, an alternative approach has been explored using laterally coupled Bragg gratings[15–17], where the grating is positioned on both sides of the ridge waveguide. In this configuration, the feedback mechanism arises from the overlap of the evanescent part of the laterally confined modes with surrounding materials, such as epitaxial layers[15], dielectrics[16], or metals[17]. While this technique bypasses the difficulties associated with the regrowth process, it introduces other challenges related to the grating coupling. Specifically, the grating coupling coefficient (κ) for higher-order modes can easily exceed that for the fundamental mode[18], leading to mode competition and mode hopping due to spatial hole burning. This necessitates precise control over the ridge waveguide width to suppress higher-order modes, which imposes limits on the device's output power and the strength of the grating coupling. Furthermore, etching surface gratings directly onto the ridge waveguide to achieve single-mode operation requires a careful balance between grating depth, waveguide losses, and metal electrode absorption[19–21].

Inspired by the wet oxidation process in vertical-cavity surface-emitting laser (VCSEL) fabrication, where high-aluminum-content layers are oxidized into low-refractive-index and insulating aluminum oxide to simultaneously form current confinement and optical confinement windows, previous studies have attempted to integrate this structure into low-loss optical waveguides for edge-emitting lasers[22]. This approach has enabled the fabrication of low-threshold current density Fabry-Pérot (FP) lasers[23–25]. However, to our knowledge, this structure has not yet been applied to the fabrication of high-quality, stable single-mode DFB lasers.

In this paper, we introduce an alternative design that incorporates an oxidized aperture confined surface grating (OASG) structure, as shown in Figure 1a. The OASG-DFB laser comprises two key components: (i) a low-refractive-index asymmetric dual-layer $Al_2O_3$ oxidized aperture waveguide, naturally formed via wet oxidation, which confines both current and the optical mode (Figure 1b); and (ii) an etched surface Bragg grating providing feedback for single-mode operation. This structure eliminates the need for ridge waveguide etching, reducing non-radiative surface recombination losses and surface optical losses from etching-induced roughness and scattering, potentially improving differential quantum efficiency. The integration of a surface grating atop the waveguide further enhances the coupling efficiency of the fundamental mode and simultaneously addresses the issue of instability in the sidewall grating's κ. This instability, caused by uncertainties in conventional etching processes, can lead to mode hopping and reduced performance. Instead, the OASG-DFB structure provides a more stable and efficient platform for the fabrication of high-performance QD DFB lasers.

Our experimental results demonstrate the robust performance of the OASG-DFB laser, which exhibits stable single-mode operation over a broad range of operating currents. The laser reaches a maximum output power of 15.1 mW, and a side-mode suppression ratio (SMSR) of up to 44 dB. Across a 4-channel array, the wavelength interval is precisely controllable at 3.5 ± 0.2 nm. The laser's linewidth is as narrow as 1.79 MHz, indicating high quality and low noise. These results highlight the OASG-DFB laser's potential as an effective solution for high-quality single-mode laser sources on the InAs/GaAs QD platform.

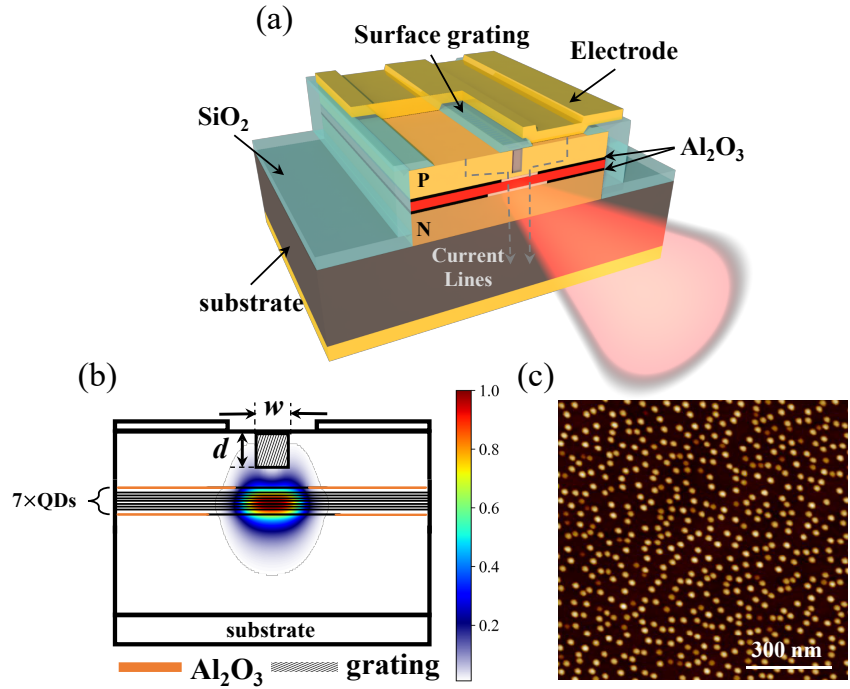

Figure 1. (a) Schematic of the OASG-DFB laser design, illustrating the structural components including the oxidized aperture waveguide and surface grating. (b) Distribution of the fundamental TE mode supported by the oxidized aperture waveguide. (c) The atomic force microscopy (AFM) image of an uncapped QD layer.

The epitaxial structure of the QD OASG-DFB laser was grown on a 3-inch GaAs (001) substrate using a solid-source molecular beam epitaxy (MBE) system. The structure starts with a 500 nm Si-doped GaAs buffer layer, followed by a 1.8 μm $Al_{0.4}Ga_{0.6}As$ n-cladding layer. A 30 nm high-Al-content $Al_{0.97}Ga_{0.03}As$ oxidation layer was then deposited. To ensure smooth band alignment and reduce electrical resistance, a 40 nm composition-graded transition layer ($Al_{0.4-0.8}Ga_{0.6-0.2}As$) was inserted between the n-cladding and oxidation layers. The active region comprises seven layers of p-doped InAs/GaAs QDs with a density of $5.5×10^{10}$ $cm^{-2}$, as determined by atomic force microscopy (AFM) measurements on uncapped QD samples (Figure 1c). P-type oxidation and transition layers were used to achieve an up and down optical confinement. A 1000 nm Be-doped $Al_{0.4}Ga_{0.6}As$ p-cladding layer and a 200 nm GaAs contact layer were then deposited. The reduced thickness of the p-cladding layer was specifically designed to optimize the coupling strength between the mode and the surface grating, while minimizing free carrier absorption losses in the heavily doped p-type contact layers. This design also helps to reduce the series resistance of the laser diode (detailed in Supplementary Information Fig. S1).

For device fabrication, approximately 300 nm of $SiO_2$ was first deposited on the wafer surface using inductively coupled plasma chemical vapor deposition (ICP-CVD) to serve as a hard mask. The third-order grating and alignment markers were defined using electron-beam lithography (EBL). After patterning, a fluorine-based reactive ion etching (RIE) process was applied to transfer the pattern into the $SiO_2$ layer. This was followed by chlorine-based inductively coupled plasma reactive ion etching (ICP-RIE) to etch the GaAs layer to a depth of approximately 1 μm, forming a uniformly periodic surface grating structure (Figure 2a). After removing the remaining $SiO_2$, a mesa was etched through the epi-stack to expose the $Al_{0.97}Ga_{0.03}As$ oxidation layers at its edge. The mesa was carefully aligned with the grating's center to ensure that the oxidation fronts progress symmetrically with the grating position, facilitating symmetric current injection. Subsequently, the $Al_{0.97}Ga_{0.03}As$ layers were laterally oxidized from the mesa edge toward the grating center at 400°C in a $H_2O/N_2$ ambient for 30 minutes, forming an insulating, low-refractive-index aluminum oxide layer. The detailed transmission electron microscopy (TEM) structure of the oxidized layers and active region is presented in Figure 2b. Furthermore, we performed energy-dispersive spectroscopy (EDS) analysis to confirm the compositional changes after oxidation (Supplementary Information Fig. S3). Photoluminescence (PL) measurements before and after oxidation indicate that the QD performance remained unaffected, confirming that the oxidation process did not introduce significant degradation (Supplementary Information Fig. S4). Our experiments showed that the oxidation rates of the p-side and n-side were approximately 150 nm/min and 110 nm/min, respectively, resulting in a vertically asymmetric oxidation waveguide biased toward the n-side, which is beneficial for wider single-mode oxidized aperture width and reducing free carrier absorption loss (detailed in Supplementary Information Fig. S1). A cross-sectional scanning electron microscopy (SEM) image of the final device is shown in Figure 2c.

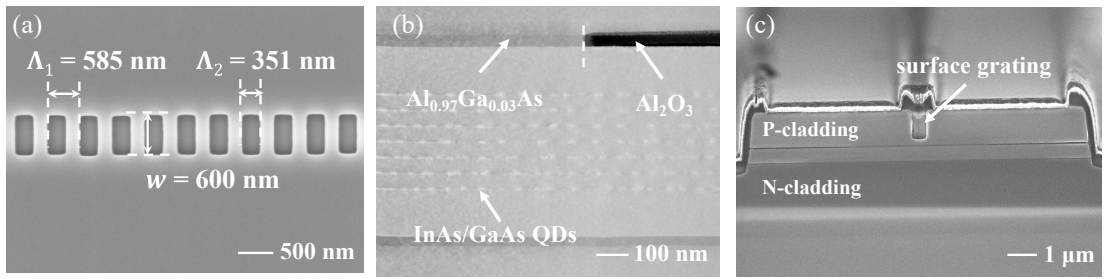

Figure 2. (a) SEM image of the fabricated OASG-DFB laser, with the white dashed line indicating the simulated current injection cross-section. (b) TEM image of the double oxide layers along with the active region. (c) SEM image of third-order surface grating.

To determine an appropriate grating coupling coefficient, two-dimensional finite-difference eigenmode (FDE) simulations were conducted to optimize the grating etch depth ($d$) and width ($w$), as shown in Figure 3a. Our results indicate that the maximum achievable κ is approximately 3 mm$^{-1}$, and for cavity lengths greater than 400 nm, we can maintain κL within the range of 1-2, ensuring optimal mode selectivity and minimal optical loss[26]. Ultimately, we selected a 1.5 mm cavity length with $d$ = 1.0 μm and $w$ = 600 nm, which corresponds to an estimated coupling constant of κL ≈ 1.65 using coupled mode theory[27]. The aperture width, which significantly influences current injection distribution in the active region, was further optimized through simulation. A 4 μm-wide current monitor was placed at the center of the active layer to evaluate the impact of different oxidized aperture widths on the current injection profile. As shown in Figure 3b, excessively wide apertures result in reduced central injection efficiency, lowering the fundamental modal gain. By controlling the mesa width and oxidation time, we achieved aperture widths of 2-3 μm, ensuring optimal interaction between the fundamental mode and the injected current.

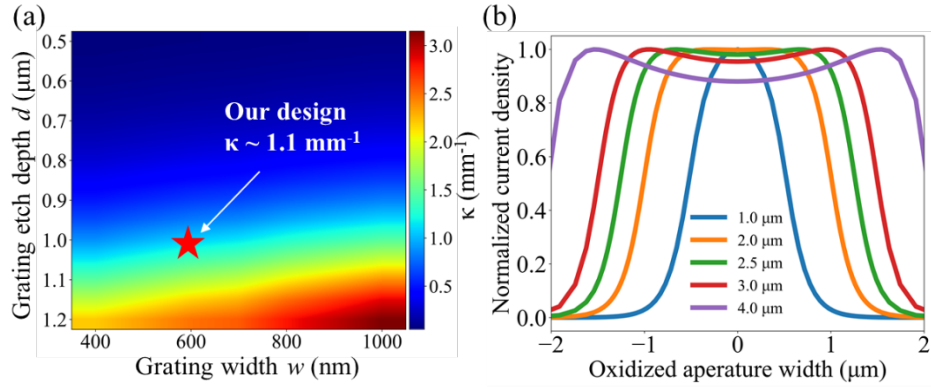

Figure 3. (a) Simulation of coupling strength as a function of surface grating etch depth and width. (b) Normalized current injection density in the oxidized aperture as a function of aperture width.

Figure 4a presents the typical *L-I-V* characteristics of the 3.0 μm × 1.5 mm device, measured under continuous-wave (CW) conditions at 25°C. The device achieves a maximum output power of 15.1 mW at 180 mA. Additionally, the characteristic temperature of the device is approximately 74 K (Supplementary Information IV). The threshold current, determined from the first derivative of the *L-I* curve, is approximately 56.8 mA, corresponding to a threshold current density of 1262 A/cm². The slightly elevated threshold current density, compared to the 571 A/cm² observed in FP lasers with the same epitaxial structure (Supplementary Information Figure S6), can be attributed to the relatively low coupling coefficient and the high turn-on voltage (~2.2 V). The inclusion of the high-Al-content oxidation layer introduces an energy barrier for both electrons and holes before reaching the active region, resulting in high resistance and reduced injection efficiency below threshold (Supplementary Information Fig. S7). However, above threshold, the differential resistance decreases to ~13 Ω(Figure 4b), which is comparable to previous reports on QD DFB devices(~10 Ω)[16], indicating that the oxidation layer does not significantly increase the resistance of the p-i-n junction. Further optimization of the oxidation layer's thickness and doping levels is expected to reduce the turn-on voltage and improve overall device performance.

Nevertheless, the device demonstrates excellent single-mode performance, as shown in Figure 5a. Stable single-mode operation is maintained across the entire current range (100-200 mA). At an injection current of 160 mA (2.82×$I_{th}$), the device achieves a maximum side-mode suppression ratio (SMSR) of 44 dB. Additionally, the device exhibits a fast wavelength shift with current, with a drift rate of 0.0186 nm/mA. This wavelength shift is primarily attributed to the conversion of electrical input power into thermal energy, which increases the device temperature, as detailed in Supplementary Information (Figure S8). Beyond individual devices, we also fabricated a four-wavelength laser array with a precise wavelength spacing of 3.5 ± 0.2 nm, ranging from 1282.38 nm to 1292.02 nm, by adjusting the grating period from 578.9 nm to 584.9 nm, as shown in Figures 5b and 5c, which corresponds to an effective mode refractive index of 3.323. The distinctive combination of the oxidized aperture waveguide structure and surface-etched grating provides an effective approach for realizing multi-wavelength, highly stable, single-mode QD-DFB lasers, which show great potential for applications in local area network wavelength division multiplexing (LWDM) and dense wavelength division multiplexing (DWDM) systems within the O-band. Furthermore, we used the delayed self-heterodyne method[28,29] to measure the Lorentzian linewidth of our OASG-DFB laser, as shown in Figure 6. The data were fitted with a Voigt profile to separate the Gaussian broadening due to 1/f noise from the long delay fiber, allowing precise determination of the Lorentzian linewidth. At a bias current of 100 mA and a temperature of 25°C, the fitted results revealed a minimum Lorentzian linewidth of approximately 1.79 MHz with a fitting confidence of 98.7%, confirming the narrow-linewidth characteristics of this device. The narrow linewidth of our device can be attributed to the three-dimensional carrier confinement of QDs, which results in a larger energy level separation and reduces carrier-induced phase noise, thereby minimizing linewidth broadening. Additionally, the oxidized aperture waveguide and surface-etched grating contribute to improved mode confinement and reduced optical losses, further suppressing linewidth broadening.

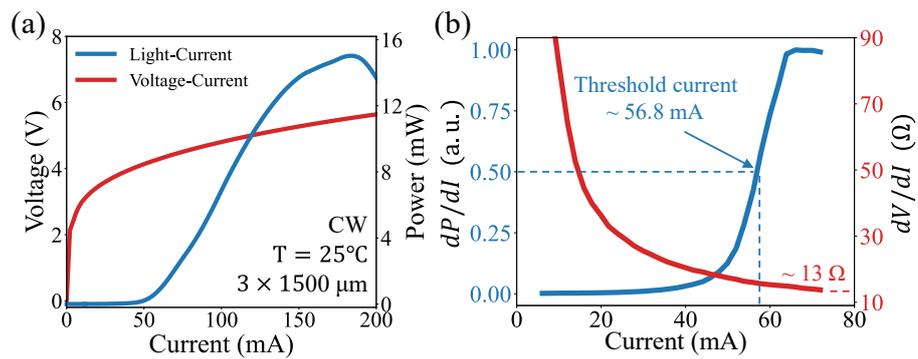

Figure 4. (a) Light-current-voltage (L-I-V) characteristics of the OASG-DFB laser under continuous-wave (CW) operation at 25°C. (b) Normalized light output power along with the derivative of current and differential resistance.

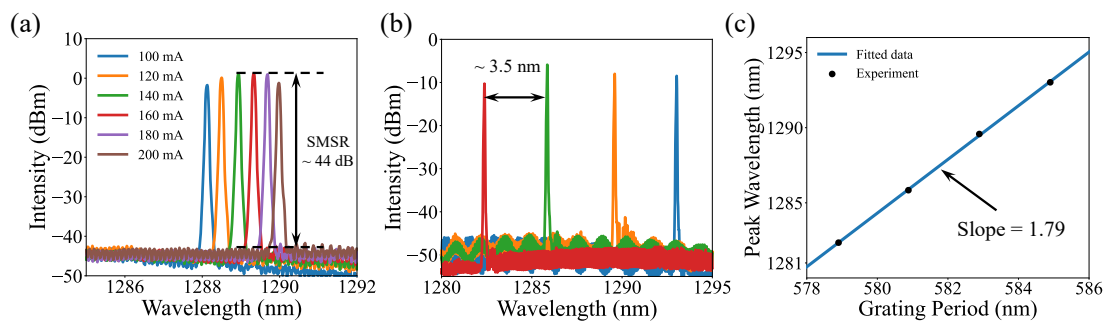

Figure 5. (a) Emission spectra of the OASG-DFB laser at different injection currents at 25°C. (b) Lasing spectra of a multi-channel array measured at an injection current of 90 mA. (c) Measured and fitted multi-wavelength laser array as a function of grating period, showing precise wavelength spacing.

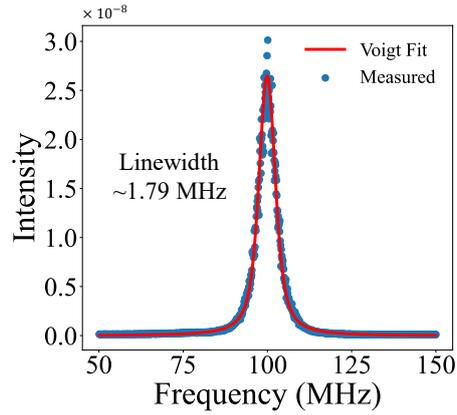

Figure 6. Radio-frequency (RF) spectrum and Voigt-profile fitted data of the OASG-DFB laser measured at 100 mA under 25°C.

In summary, we have developed a fabrication approach for QD DFB lasers with high single-mode stability, using the integration of a dual-layer oxidized aperture as a low-refractive-index confinement waveguide and efficient fundamental mode coupling achieved through a third-order surface grating. This design streamlines the fabrication process, eliminates the need for regrowth, and significantly reduces the challenges associated with etching, offering a robust and scalable method for producing high-performance QD DFB lasers. While the oxidation layer introduces some electrical resistance, this structural feature is a proven component in commercial VCSELs. With further optimization of the oxidation layer's thickness, doping concentration, and epitaxial placement, this limitation can be effectively addressed. Our fabrication approach not only simplifies manufacturing and lowers production costs but also enhances the versatility of GaAs-based QD DFB laser production. This method demonstrates exceptional potential for both single-die fabrication and heterogeneous integration, paving the way for next-generation photonic devices.

**Supplementary Material**

The supplementary material provides detailed information on the characterization of the laser wafer, including PL spectroscopy and SEM imaging, which confirm the high quality of the QD active layer and the uniformity of the fabricated structures. It also includes a comprehensive discussion on the fabrication process, including the wet oxidation procedure and the resulting asymmetric oxidation waveguide, supported by infrared microscopy images. Additionally, the material presents simulations of the optical field overlap in the p-contact layer, highlighting the influence of the p-cladding thickness on optical losses. Further, it contains a detailed analysis of the device's thermal resistance, with measurements taken under continuous-wave and pulsed conditions, and discusses the impact of thermal effects on the device's performance. These data provide essential insights into the optimization of our QD OASG-DFB laser design. Finally, we have included a table comparing the performance of this device with QD-DFB lasers reported recently.

**Acknowledgements**

This work was supported by National Natural Science Foundation of China (62422516) and National Key R&D Program of Guang-dong Province (2020B0303020001).

**Conflict of Interest**

The authors have no conflicts to disclose.

**Corresponding Author**

*Corresponding author: yuying26@mail.sysu.edu.cn

**Author Contributions**

**Zhengqing Ding:** Conceptualization (equal); Data curation (equal); Formal analysis (main); Investigation (equal); Methodology (equal); Software (main); Validation (equal); Visualization (main); Writing – original draft (main); Writing – review & editing (equal). **Anyao Zhu:** Data curation (equal); Formal analysis (equal); Investigation (equal); Writing – review & editing (equal). **Chaoyuan Yang:** Investigation (equal); Methodology (equal). **Kun Zhan:** Investigation (equal); Methodology (equal). **Yingxin Chen:** Investigation (equal); Methodology (equal). **Ying Yu:** Conceptualization (main); Funding acquisition (main); Investigation (equal); Methodology (main); Project administration (main); Supervision (main); Validation (equal); Writing – original draft (equal); Writing – review & editing (main). **Siyuan Yu:** Methodology (equal); Validation (equal); Writing – review & editing (equal).

**Data availability**

The data that support the findings of this study are available from the corresponding author upon reasonable request.